\documentclass[aps,pra,twocolumn,showpacs,amsmath,amssymb]{revtex4}

\usepackage{graphicx} 
\usepackage{dcolumn} 
\usepackage{bm} 
\usepackage{epsfig}
\usepackage{amsmath}
\usepackage{amssymb}
\usepackage{hyperref}
\usepackage{float}
\usepackage{color}
\usepackage{ulem}
\begin{document}
\newcommand{\bsy}[1]{\mbox{${\boldsymbol #1}$}} 
\newcommand{\bvecsy}[1]{\mbox{$\vec{\boldsymbol #1}$}} 
\newcommand{\bvec}[1]{\mbox{$\vec{\mathbf #1}$}} 
\newcommand{\btensorsy}[1]{\mbox{$\tensor{\boldsymbol #1}$}} 
\newcommand{\btensor}[1]{\mbox{$\tensor{\mathbf #1}$}} 
\newcommand{\tensorId}{\mbox{$\tensor{\mathbb{\mathbf I}}$}} 
\newcommand{\be}{\begin{equation}}
\newcommand{\ee}{\end{equation}}
\newcommand{\bea}{\begin{eqnarray}}
\newcommand{\eea}{\end{eqnarray}}
\newcommand{\e}{\mathrm{e}}
\newcommand{\arccot}{\mathrm{arccot}}
\newcommand{\arctanh}{\mathrm{arctanh}}

\title{Electromagnetic propagation in a relativistic electron gas at finite temperatures}

\author{D. M. Reis$^{1}$, E. Reyes-G\'omez$^{2}$, L. E. Oliveira$^{3}$, and C. A. A. de Carvalho$^{4,5}$}

\affiliation{
$^1$Centro Brasileiro de Pesquisas F\'{\i}sicas - CBPF, Rua Dr. Xavier Sigaud 150, Rio de Janeiro - RJ, 22290-180, Brazil\\
$^2$Instituto de F\'{i}sica, Universidad de Antioquia - UdeA, Calle 70 No. 52-21, Medell\'{\i}n, Colombia\\
$^3$Instituto de F\'{i}sica, Universidade Estadual de Campinas - Unicamp, Campinas - SP, 13083-859, Brazil \\
$^4$Diretoria Geral de Desenvolvimento Nuclear e Tecnol\'{o}gico da Marinha - DGDNTM, Rua da Ponte s/n, Ed. 23 do AMRJ, Rio de Janeiro - RJ, 20091-000, Brazil\\
$^5$Instituto de F\'{\i}sica, Universidade Federal do Rio de Janeiro - UFRJ, Caixa Postal  68528, Rio de Janeiro - RJ, 21945-972, Brazil}

\begin{abstract}
We describe electromagnetic propagation in a relativistic electron gas at finite temperatures and carrier densities. Using quantum electrodynamics at finite temperatures, we obtain electric and magnetic responses and general constitutive relations. Rewriting the propagator for the electromagnetic field in terms of the electric and magnetic responses, we identify the modes that propagate in the gas. As expected, we obtain the usual collective excitations, i.e., a  longitudinal electric and two transverse magnetic plasmonic modes. In addition, we find a purely photonic mode that satisfies the wave equation in vacuum, for which the electron gas is transparent. We present dispersion relations for the plasmon modes at zero and finite temperatures and identify the intervals of frequency and wavelength where both electric and magnetic responses are simultaneously negative, a behavior previously thought not to occur in natural systems.
\end{abstract}


\date{\today}

\maketitle

\section{Introduction}

The knowledge of the responses of a relativistic electron gas (REG) at finite temperatures and densities to electromagnetic (EM) radiation is a useful tool for understanding the physics of several systems, as we will outline below. The present article uses a quantum field theory treatment to describe the interaction of the EM field with the REG. It focuses on the calculation of the responses and on the description of the modes of propagation within the gas.

Investigations of the response of a REG to the action of an external electromagnetic (EM) perturbation have many similarities with studies in the fields of photonics and plasmonics \cite{yan2015hydrodynamic, govorov2014photogeneration}, where it is crucial to understand the propagation of EM and plasmonic modes that are also present in the REG. 

In photonics and plasmonics one normally uses phenomenological expressions for the responses of the media of interest. Here, however, one may actually compute such responses from first-principles, so that we envisage applying our techniques in the future to the actual computation of the responses of artificially constructed materials. 

In the context of plasma physics, the REG has been the subject of many articles \cite{plasma refs}. The treatment presented here  uses a field theory approach that can be quite naturally generalized to encompass the study of several other plasmas, whose physical properties may be then compared to those of the REG.

In astrophysical scenarios, it has been shown that electron-positron pairs may be obtained in systems such as compact stars \cite{han2012electron, berezhiani2015}. There again, the study of the REG is of importance, especially with respect to the propagation of EM and collective modes through the gas, which may even show up in the observations.

Related situations of interest whose analysis can profit from a good understanding of the REG may be obtained in the laboratory by using high intensity lasers in a plasma gas \cite{Nerush2011}, with the laser photons acquiring an effective mass in the medium, thus characterizing a collective excitation \cite{raicher2014, raicher2016}. 

Finally, heavy-ion collison experiments at RHIC and LHC \cite{Aamodt2010} produce collective excitations in the hot QCD liquid matter, known as the quark-gluon plasma (QGP). Its transverse and longitudinal collective plasma excitations, also known as plasmons, have similar behaviour to the plasmons in a REG \cite{Yousuf2017}.  

Recently, it has been theorized \cite{AragaoPRDS2016} that the REG might be a candidate for the realization of a natural metamaterial, understood, as proposed originally by Veselago in 1968 \cite{Veselago}, as  a material that can have simultaneously negative values for its electric permittivity and magnetic permeability. 

Although the existence of natural metamaterials has not been reported so far, such systems have already been artificially constructed. A successful design of an artificial metamaterial was reported by Smith {\it et al} \cite{Smith} and consisted of a periodic array of metallic split-ring resonators and wires, which exhibited simultaneously negative values of the  electric permittivity ($\epsilon$) and  magnetic permeability ($\mu$) in a microwave-frequency band of the EM spectrum.  Subsequently, metamaterials were realized up to the visible range \cite{Zhe} and all-dielectric metamaterials based on Si/SiO$_2$ heterostructures have been reported \cite{Moitra}. 

The possibility that the REG may behave as a natural metamaterial \cite{AragaoPRDS2016} deserves further investigation. To investigate the EM responses of the REG it is natural to adopt  the quantum-electrodynamics (QED) formalism for physical systems at finite temperatures  \cite{Kapusta, AP}, within linear response and in the random-phase approximation (RPA). 

In that formalism, it has been shown  that, in the limit of temperature $T \rightarrow 0$, theoretical results for the electric permittivity agree with previous non-relativistic calculations performed by Lindhard \cite{Lindhard1954}. Moreover, in the long-wavelength limit both $\epsilon$ and $\mu^{-1}$ exhibit Drude-like EM responses and may be simultaneously negative \cite{AragaoPRDS2016}. 

In addition, the validity of the model has been tested in the non-relativistic limit  by successfully \cite{ReyezEPL2016} describing the experimental behavior of the plasmon energy, as a function of both temperature and wave vector, in low-energy condensed-matter systems such as graphite \cite{Jensen,Portail} and tin oxide \cite{Zou}. The study of the EM response of a REG in the regime of high temperatures and densities is, therefore, in order.

The aim of the present work is to investigate the behavior of the EM response functions of the REG, as functions of the temperature, density, frequency, and  wave vector, as well as the EM propagation modes within the REG. This study is organized as follows: In section \ref{Prop} we describe the theoretical procedure used for obtaining the EM responses and propagation in the REG. Results are presented in section \ref{results} and conclusions are in section \ref{conclusions}.

\section{Electromagnetic propagation}
\label{Prop}
In \cite{AragaoPRDS2016} and \cite{AragaoSubmit}, we have used a semiclassical expansion, where the electromagnetic part is treated as a classical external field plus quantum fluctuations, but the electrons are subject to a full quantum treatment, just as it is done in the nonrelativistic case that leads to the celebrated Lindhard expression \cite{Lindhard1954}.

In fact, integrating over the electron field yields a fermionic determinant which is expanded in the classical gauge field, assumed to be weak. Only the lowest-order term in the expansion is kept, which is equivalent to the linear response approximation (terms of order $\alpha(\alpha E^2/m^4)$ or $\alpha(\alpha B^2/m^4)$  are neglected), and electron–electron interactions are also neglected, as they give rise to higher-order terms in $\alpha \equiv e^2/(4\pi) = 1/137$. The responses we compute are quantum mechanical, only the electromagnetic fields are treated classically.

The partition function for QED at finite temperatures in linear response and RPA is given by a quadratic functional integral in Euclidean space over fields obeying $A_\mu (\beta, \vec{x})=A_\mu (0, \vec{x})$ \cite{AragaoPRDS2016, Kapusta} 
\be
\label{quadpart}
Z[A]= \oint [dA_\mu] \det[-\partial ^2] \exp ( -\frac {1}{2} \sum \!\!\!\!\! \!\!\!\int {\tilde A}_\mu {\tilde \Gamma}_{\mu \nu} {\tilde A}_\nu),
\ee
\be
\label{Gamma}
{\tilde \Gamma}_{\mu \nu}= q^2 \delta_{\mu\nu} - (1- \frac{1}{\lambda}) q_\mu q_\nu - {\tilde \Pi}_{\mu \nu}.
\ee
$q^2= q^2_4+|\vec{q}|^2$ and $\tilde{A}_\mu$ is the Fourier transform of the gauge field $A_\mu$.
The determinant comes from the Lorentz gauge condition and $\lambda$ is a gauge parameter. We have used the simplified notation 
\be
\sum \!\!\!\!\! \!\!\!\int \equiv \frac {1}{\beta} \!\! \sum_{n=-\infty} ^ {+\infty}\! \!\int\! \frac {d^3 q}{(2\pi)^3},
\ee
and the polarization tensor of QED \cite{AP}
\be
\tilde{\Pi}_{\mu\nu} = - \frac {e^2}{\beta} \! \sum_{n=-\infty} ^ {+\infty}\! \int\! \frac {d^3 p}{(2\pi)^3} \hbox{Sp} [\gamma_\mu G_0(p) \gamma_\nu G_0 (p-q)],
\label{pol}
\ee
where $G_0$ is the free electron propagator at finite density. The summation over $n$ in \eqref{pol} is performed over Matsubara frequencies $p_4=(2n+1)\pi T$.

One may write $\tilde{\Pi}_{\nu\sigma}=\tilde{\Pi}_{\nu\sigma}^{(v)} + \tilde{\Pi}_{\nu\sigma}^{(m)}$ to separate vacuum and medium contributions. The vacuum contribution is 
\be
 -\frac{\tilde{\Pi}_{\nu\sigma} ^{(v)}}{ q^2 }= \left (\delta_{\nu\sigma} - \frac{q_\nu q_\sigma}{q^2}\right ) {\cal{C}}(q^2),
\label{vac}
\ee
whereas the medium contribution is given by
\bea
\label{med1}
&& -\frac{\tilde{\Pi}_{ij} ^{(m)}}{ q^2 }= \left (\delta_{ij} - \frac{q_iq_j}{|\vec{q}|^2}\right ) {\cal{A}} + \delta_{ij} \frac{q_4^2}{|\vec{q}|^2} {\cal{B}},  \\
&& -\frac{\tilde{\Pi}_{44} ^{(m)}}{q^2 } = {\cal B},  \,\,\,\,\,\,\,\,\,  -\frac{\tilde{\Pi}_{4i} ^{(m)}}{q^2 } = - \frac{q_4 q_i}{|\vec{q}|^2} {\cal B},
\label{med2}
\eea
where the three scalar functions ${\cal{A}} (q_4, |\vec{q}|)$, ${\cal B} (q_4, |\vec{q}|)$, and ${\cal{C}}(q^2 = q_4^2+ |\vec{q}|^2)$ are determined from the Feynman graph in Eq. \eqref{pol} \cite{AragaoPRDS2016}.

Following \cite{Kapusta}, we introduce the  projector 
\be
\label{proj}
{\cal P}_{\mu\nu}= \delta_{\mu\nu} - \frac{q_\mu q_\nu}{q^2},
\ee 
and the transverse ${\cal P}^T_{\mu\nu}$ projector
\bea
\label{projT}
&& {\cal P}^T_{ij}= \delta_{ij} - {\hat q}_i {\hat q}_j,\\
&& {\cal P}^T_{44}={\cal P}^T_{4i}=0,
\eea
with ${\hat q}_i = q_i/|{\vec q}|$. The longitudinal projector is then ${\cal P}^L_{\mu\nu} \equiv {\cal P}_{\mu\nu} - {\cal P}^T_{\mu\nu}$ so that the polarization tensor is given by
\be
{\tilde \Pi}_{\mu \nu}={\tilde \Pi}^{(v)}_{\mu \nu} + {\tilde \Pi}^{(m)}_{\mu \nu} = {\cal F} {\cal P}^L_{\mu\nu} + {\cal G} {\cal P}^T_{\mu\nu},
\ee
where 
\bea
\label{F}
&& {\cal F} = -q^2 \left ({\cal C}+ {\cal B} + \frac{q_4^2}{|\vec q |^2} {\cal B}\right ), \\
&& {\cal G} = -q^2 \left ({\cal C}+ {\cal A} + \frac{q_4^2}{|\vec q |^2} {\cal B}\right ).
\label{G}
\eea
The quadratic kernel is then
\be
\label{kernelgamma}
{\tilde \Gamma}_{\mu \nu} = (q^2 - {\cal F}) {\cal P}^L_{\mu\nu} +  (q^2 - {\cal G}) {\cal P}^T_{\mu\nu} + \frac{1}{\lambda} q_\mu q_\nu,
\ee
and its inverse, the photon propagator, reads
\be
\label{photprop}
{\tilde \Gamma}^{-1}_{\mu \nu} = \frac{ {\cal P}^L_{\mu\nu}}{q^2 - {\cal F}} + \frac{ {\cal P}^T_{\mu\nu}}{q^2 - {\cal G}} + \frac{\lambda}{q^2} \frac{q_\mu q_\nu}{q^2}.
\ee

 On the other hand, one may also obtain EM responses from the polarization tensor. Indeed, we have recently \cite{AragaoPRDS2016} shown  that the (Fourier transformed) constitutive equations of the REG are  $D_j=\epsilon_{jk} E_k + \tau_{jk}  B_k$, $ H_j= \nu_{jk}  B_k + \tau_{jk}  E_k$, where we have used the notation $\nu\equiv \mu^{-1}$. The linear-response RPA tensors are $\epsilon_{jk}= \epsilon \delta_{jk} + \epsilon' \hat{q}_j \hat{q}_k$, $\nu_{jk}= \nu \delta_{jk} + \nu' \hat{q}_j \hat{q}_k$, and $ \tau_{jk}= \tau \epsilon_{jkl} \hat{q}_l$. 
The eigenvalues of $\epsilon_{jk}$ are $\epsilon + \epsilon'$ and $\epsilon$. The eigenvector associated to $\epsilon+ \epsilon'$ is along $ \hat{q}_k$, thus longitudinal, whereas the two eigenvectors corresponding to the eigenvalues $\epsilon$ are in directions transverse to $ \hat{q}_k$. The same occurs for $\nu_{jk}$, with eigenvalues $\nu + \nu '$ and $\nu$, whereas $\tau_{jk}$ is clearly transverse.  

The permittivities and inverse permeabilities are determined by the three scalar functions  ${\cal A}^\ast$, ${\cal B}^\ast$ and ${\cal C}^\ast$, with the asterisk denoting the continuation to Minkowski space, $q_ 4 \rightarrow i\omega - 0^+$, of the Euclidean scalar functions ${\cal{A}} (q_4, |\vec{q}|)$, ${\cal B} (q_4, |\vec{q}|)$ and ${\cal{C}}(q_4^2 + |\vec{q}|^2)$ obtained from the polarization tensor $\tilde \Pi_{\mu\nu}$ \cite{AragaoPRDS2016, corr, IZ}, i.e., 
\bea
\label{e6}
\epsilon \!\!&=& \!\!1 + {\cal A^*}  + \left ( 1 - \frac{\tilde{\omega}^2}{\tilde{q}^2} \right ) {\cal B^*}
+ \left ( 2 + \frac{\tilde{\omega}^2}{\tilde{q}^2 - \tilde{\omega}^2} \right ) {\cal C^*},
\eea
\bea
\label{e7}
\nu  &=& 1 + {\cal A^*}  - 2 \frac{\tilde{\omega}^2}{\tilde{q}^2} {\cal B^*} 
+ \left ( 2 - \frac{\tilde{q}^2}{\tilde{q}^2 - \tilde{\omega}^2} \right ) {\cal C^*},
\eea
\bea
\label{e8}
\epsilon^{\prime} &=& - {\nu^{\prime}}  = - \left [ {\cal A^*} + \frac{\tilde{q}^2}{\tilde{q}^2 - \tilde{\omega}^2} {\cal C^*}\right ]  ,
\eea
and
\bea
\label{e9}
\tau  \! &=& \! - \frac{\tilde{\omega}}{\tilde{q}} \! \! \left [ \!  \frac{\tilde{q}^2}{\tilde{q}^2 - \tilde{\omega}^2} {\cal C^* + {\cal B^*} }  \right ] \! \!.
\eea 
For the longitudinal responses, one then obtains
\bea
\label{long1}
&& \epsilon_L = \epsilon+\epsilon'= 1+{\cal C}^\ast + \left(1-\frac{\tilde{\omega}^2}{\tilde{q}^2}\right) {\cal B}^\ast, \\
\label{long2}
&& \nu_L = \nu+\nu'=1+2{\cal C} ^\ast+ 2{\cal A}^\ast -2 \frac{\tilde{\omega}^2}{\tilde{q}^2} {\cal B}^\ast.
\eea
Our formulae make use of the dimensionless variables  $\tilde{q} = |\vec{q}| / q_c$, $\tilde{\omega} = \omega / \omega_c$, $\tilde{\beta} = m c^2 \beta$, and $\tilde{\xi} = \xi/ (m c^2)$, where  $q_c = m c / \hbar$ is the Compton wave vector, $\omega_c = m c^2 / \hbar$ is the Compton frequency, $\beta = 1 / (k_B T)$, $T$ is the absolute temperature, and $\xi$ is the chemical potential of the electron gas.

We now use Eqs. \eqref{F}, \eqref{G} and write the propagators in Minkowski space by letting $q_4 \rightarrow i \omega - 0^+$, $q^2 = q_4^2+|\vec{q}|^2 \rightarrow - q^2= |\vec{q}|^2 - \omega^2$ and $\cal A, \cal B, \cal C \rightarrow {\cal A^*}, {\cal B^*}, {\cal C^*} $. Then, one obtains
\bea
\label{poleE}
&& \frac{1}{q^2 - {\cal F}} \rightarrow \frac{1}{-q^2 \epsilon_L }, \\
&& \frac{1}{q^2 - {\cal G}} \rightarrow \frac{2}{-q^2 [\nu_L + 1]}.
\label{poleM}
\eea
Eq. \eqref{poleE} leads to a pole in the ${\cal P}^L_{\mu\nu}$ longitudinal propagator whenever 
\be
\label{eplasmon} 
\epsilon_L (\omega, |{\vec q}|) = 0.
\ee
Note that the pole at $q^2=0$ is not realized in this case as it corresponds to a transverse mode, as we shall explicitly show. We remark that  \eqref{eplasmon} corresponds to the usual condensed matter dispersion relation of longitudinal plasmon collective excitations. 

The ${\cal P}^T_{\mu\nu}$ transverse propagator [cf. Eq.  \eqref{poleM}] has poles whenever
\bea
\label{mplasmon} 
&& \nu_L (\omega, |{\vec q}|) = -1, \\
&& q^2= \omega^2 - |{\vec q}|^2 = 0.
\label{photon}
\eea
Analogously to the longitudinal case,  Eq. \eqref{mplasmon} yields the dispersion relation of transverse plasmon collective excitations whereas Eq. \eqref{photon} yields a photonic mode that propagates with the speed of light $c\, (=1)$ in vacuum.

In order to have more explicit expressions for the plasmon modes, it is useful to write the projectors as
\bea
&& {\cal P}_{\mu\nu} = n^{(1)}_\mu n^{(1)}_\nu + n^{(2)}_\mu n^{(2)}_\nu + n^{(3)}_\mu n^{(3)}_\nu , \\
&& {\cal P}^T_{\mu\nu}=n^{(1)}_\mu n^{(1)}_\nu + n^{(2)}_\mu n^{(2)}_\nu,
\eea 
where $n^{(i)}_\mu= (0, {\hat n}^{(i)})$, ${\hat q} . {\hat n}^{(i)}=0$, $|{\hat n}^{(i)}|=1$, for $i=1,2$,  satisfying ${\hat n}^{(1)}_i{\hat n}^{(1)}_j + {\hat n}^{(2)}_i{\hat n}^{(2)}_j + {\hat q}_i{\hat q}_j = \delta_{ij}$. For $n^{(3)}_\mu$, we find
\be
n^{(3)}_\mu = \left (\frac{-|{\vec q}|}{\sqrt{q^2}}, \frac{q_4 {\hat q}}{\sqrt{q^2}}\right ),
\ee 
if we demand that it must be normalized and orthogonal to $q_\mu$ and $n^{(i)}_\mu , i=1,2$, thus satisfying $n^{(1)}_\mu n^{(1)}_\nu + n^{(2)}_\mu n^{(2)}_\nu + n^{(3)}_\mu n^{(3)}_\nu +( q_\mu q_\nu / q^2)= \delta_{\mu \nu}$. Then
\bea
&& {\cal P}^L_{\mu\nu} = n^{(3)}_\mu n^{(3)}_\nu , \\
&& {\cal P}^T_{\mu\nu}=n^{(1)}_\mu n^{(1)}_\nu + n^{(2)}_\mu n^{(2)}_\nu.
\eea 
A few observations are in order: \\ \\
(i) in Minkowski space, we have
\be
n^{(3)}_\mu = \left (\frac{i |{\vec q}|}{\sqrt{q^2}}, \frac{\omega {\hat q}}{\sqrt{q^2}}\right ),
\ee 
which in the long-wavelength limit becomes $n^{(3)}_\mu= (0, {\hat q})$; \\ \\
(ii) in that limit, Ref. \cite{AragaoPRDS2016} obtains Drude expressions for $\epsilon_L= 1- (\omega_e^2/\omega^2)$ and $\nu_L=1- (\omega_m^2/\omega^2)$. Inserting this into \eqref{poleE} and \eqref{poleM}, and using the fact that $\omega_m^2= 2\omega_e^2$, we find $\omega_e^2 - \omega^2$ as the denominator for both longitudinal and transverse plasmon propagators.

The collective plasmon excitations correspond to charge density and current density oscillations. Indeed, the collective field $A^L \equiv n^{(3)}_\mu {\cal P}^L_{\mu\nu} A_\nu = A_\nu n^{(3)}_\nu$, in Euclidean space, is given by
\be
A^L=\frac{ -i {\vec q} . ( -i {\vec q} A_4 + iq_4 \vec{A})}{\sqrt{q^2} |{\vec q}|}  = \frac{- \vec{\nabla} . {\vec E}}{\sqrt{q^2} |{\vec q}|}.
\ee
Since the field has a longitudinal component, we may define an effectice charge density ${\rho_e}$ as $\vec{\nabla} . {\vec E} \equiv \rho_e (q)$. Similarly, the collective field $A^T_{\mu} \equiv {\cal P}^T_{\mu\nu} A_\nu$  is given by $(0, {\vec A}^T)$, where ${\vec A}^T= A_1 {\hat n}^{(1)} + A_2 {\hat n}^{(2)}$ and $A_i= {\vec A}. {\hat n}^{(i)}$. One then obtains
\be
{\vec A}^T=\frac{ i {\vec q} \wedge ( i {\vec q} \wedge {\vec A})}{|{\vec q}|^2}=\frac{ {\vec \nabla} \wedge {\vec B}}{|{\vec q}|^2}.
\ee
We may then define an effective current density $\vec{j}_e$ through ${\vec \nabla} \wedge {\vec B} = \vec{j}_e$. Then, if we  use \eqref{F}, \eqref{G}, and \eqref{kernelgamma}, and leave aside a gauge term, the plasmon  Lagrangean may be written, in Minkowski space, as
\be
\label{osc}
\rho_e(q)\left (\frac{\epsilon_L}{{\vec q}^2}\right )\rho_e(q) + j_{e}^k (q) \left [\frac{(\nu_L+ 1)(1-\frac{\omega^2}{|{\vec q}|^2})}{2{\vec q}^2}\right ]j_{e}^k (q),
\ee
where $q=(\omega, {\vec q})$. The above expression physically describes the interaction of charge densities induced by the longitudinal component of the fluctuating electric fields and current densities (loops in the plane perpendicular to $\hat q$) induced by the fluctuating magnetic fields. Apart from that, whenever $\epsilon_L \ne 0$ and $\nu_L \ne -1$, we just have the propagation of an electromagnetic wave with a propagator given by \eqref{poleM}.

An alternative and somewhat complementary analysis may be obtained from Maxwell's equations combined with the constitutive relations written out previously. Maxwell's equations are (we have now restored the speed of light $c$)
\begin{subequations}
\label{eq1}
\be
\label{eq1a}
q_i D_i = 0,
\ee
\be
\label{eq1b}
q_i B_i = 0,
\ee
\be
\label{eq1c}
\epsilon_{ijk} q_j E_k = \frac{\omega}{c} B_i,
\ee
and
\be
\label{eq1d}
\epsilon_{ijk} q_j H_k = - \frac{\omega}{c} D_i.
\ee
\end{subequations}
The constitutive equations were defined in the paragraph after Eq. \eqref{photprop}. From Eq. \eqref{eq1a} and the constitutive relations, we derive
\be
\label{Deq}
 ({\vec q} . {\vec E}) \epsilon_L = 0.
\ee
If $\epsilon_L \ne 0$, then we must have ${\vec q} . {\vec E} = 0$, so that  Eq. \eqref{eq1d} and the constitutive relations give
\be
\label{rotH}
\left [ \tau |{\vec q}| + \epsilon \frac{\omega}{c}\right ] {\vec E} + \left [ \nu |{\vec q}| - \tau \frac{\omega}{c}\right ] ({\hat q} \wedge {\vec B})=0,
\ee
which combined with Eq. \eqref{eq1c} yields \cite{corr2} a generalized wave equation for ${\vec E}$ (and an analogous one for ${\vec B}$) 
\be
\label{waveq}
\left [ |{\vec q}|^2 - \mu\epsilon \frac{\omega^2}{c^2} - 2 \mu\tau |{\vec q}| \frac{\omega}{c} \right ] {\vec E}=0.
\ee
However, using Eqs. \eqref{e6} to \eqref{e9}, \eqref{long1} and \eqref{long2}, Eq. \eqref{rotH} becomes
\be
\label{rotH1}
(\nu_L + 1) [{\vec q} \wedge {\vec B} + \frac{\omega}{c} {\vec E}] = 0,
\ee
whereas Eq. \eqref{waveq} yields
\be
\label{waveq1}
(\nu_L + 1) q^2 = (\nu_L + 1) \left [ \frac{\omega^2}{c^2} - |{\vec q}|^2 \right ] = 0.
\ee
We see that Maxwell's equation \eqref{eq1a} will be satisfied if $\epsilon_L=0$. This coincides with the longitudinal plasmon condition. If $\epsilon_L \ne 0$, then ${\vec E}$ is transverse and eq. \eqref{waveq1} will be satisfied if either $\nu_L = -1$ (transverse plasmons) or $q^2 = 0$ (photons). The fact that the wave equation factors out into two terms is a consequence of the specific form of the EM responses for the REG.

The plasmon modes and the photonic mode obtained from quantum responses to the electromagnetic fields will appear whenever the dispersion relation $\omega=\omega(|\vec{q}|)$ obeys one of the conditions derived on eqs. \eqref{eplasmon}-\eqref{photon} ($\epsilon_L=0$; $\nu_L=-1$ and $\omega=|\vec{q}|$, respectively). Otherwise, the electromagnetic field will propagate with responses given by $\epsilon_{ij}(\omega, |\vec{q}|)$ and $\nu_{ij}(\omega, |\vec{q}|)$.

Before proceeding, we note that ${\cal A^*}, {\cal B^*}, {\cal C^*}$ are given explicitly by \cite{AragaoPRDS2016,ReyezEPL2016}
\bea
\label{eq5}
{\cal A^*}  &=&  \frac{A_{\alpha}}{\tilde{q}^2-\tilde{\omega}^2} \, {\cal I}+ \left [ 1 - \frac{3}{2} \frac{\tilde{q}^2-\tilde{\omega}^2}{\tilde{q}^2} \right ] {\cal B^*},
\eea
\be
\label{eq6}
{\cal B^*} =  \frac{B_{\alpha}}{\tilde{q}^2-\tilde{\omega}^2} \, {\cal J} ,
\ee
and
\bea
\label{eq7}
{\cal C^*} &=& - C_{\alpha} \biggl \{ \frac{1}{3} + \left ( 3 + \gamma^2 \right )  \left [ \gamma \, \arccot (\gamma)  -1  \right ] \biggr \},
\eea
where
\be
\label{eq8}
\gamma = \sqrt{\frac{4}{\tilde{\omega}^2 - \tilde{q}^2}-1},
\ee
$A_{\alpha} = B_{\alpha} = 4 \alpha /\pi$, $C_{\alpha} = A_{\alpha} / 12$, $\alpha$ is the fine-structure constant, and the functions $\cal I$ and $\cal J$ are the one-dimensional integrals

\bea
\label{eq9}
{\cal I}  &=& \int_{0}^{\infty} dy \, \frac{y^2}{\sqrt{y^2+1}} {\cal F}_0 (y,\tilde{\beta},\tilde{\xi}) \nonumber \\ &\times& \left [ 1 + \frac{2 - \tilde{q}^2 + \tilde{\omega}^2}{8 y \tilde{q}} {\cal F}_1 (y,\tilde{q},\tilde{\omega}) \right ]
\eea

and
\bea
\label{eq10}
{\cal J}  &=& \int_{0}^{\infty} dy \, \frac{y^2}{\sqrt{y^2+1}} {\cal F}_0 (y,\tilde{\beta},\tilde{\xi}) \nonumber \\ &\times& \left [ 1 + \frac{4 (y^2+1) - \tilde{q}^2 + \tilde{\omega}^2}{8 y \tilde{q}} {\cal F}_1 (y,\tilde{q},\tilde{\omega}) \right. \nonumber \\ &-& \left. \frac{\tilde{\omega} \sqrt{y^2+1}}{2 y \tilde{q}} {\cal F}_2 (y,\tilde{q},\tilde{\omega}) \right ],
\eea
respectively. The functions ${\cal F}_0, {\cal F}_1$ and ${\cal F}_2$ are defined as

\be
\label{eq11}
{\cal F}_0 \! \left ( y,\tilde{\beta},\tilde{\xi} \right ) \! = \! \frac{1}{\e^{\tilde{\beta} \left ( \sqrt{y^2 + 1} - \tilde{\xi} \right )} \! + \! 1} \! - \! \frac{1}{\e^{\tilde{\beta} \left ( \sqrt{y^2 + 1} + \tilde{\xi} \right )} \! + \! 1},
\ee

\be
\label{eq12}
{\cal F}_1 (y,\tilde{q},\tilde{\omega}) \! = \! \ln \! \left [ \frac{(\tilde{q}^2 \! - \! \tilde{\omega}^2 + 2 y \tilde{q})^2 \! - \! 4 (y^2 + 1) \tilde{\omega}^2}{(\tilde{q}^2 \! - \! \tilde{\omega}^2 - 2 y \tilde{q})^2 \! - \! 4 (y^2 + 1) \tilde{\omega}^2} \right ] \!,
\ee
and
\be
\label{eq13}
{\cal F}_2 (y,\tilde{q},\tilde{\omega}) \! = \! \ln \! \left [ \! \frac{\tilde{\omega}^4 - 4 (\tilde{\omega} \sqrt{y^2+1} + y \tilde{q})^2}{\tilde{\omega}^4 - 4 (\tilde{\omega} \sqrt{y^2+1} - y \tilde{q})^2} \! \right ] \!.
\ee
respectively.

\section{Results and discussion}
\label{results}

\subsection{The chemical potential}

\begin{figure}
\epsfig{file=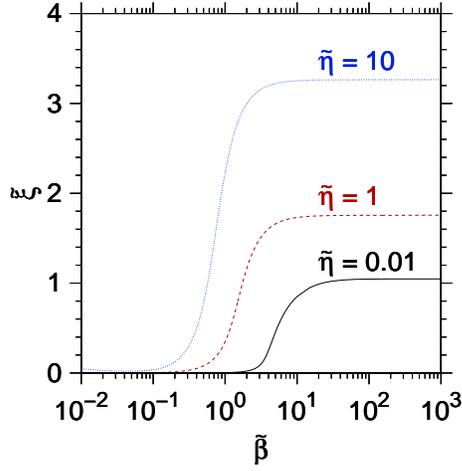,width=0.7\columnwidth}
\caption{(Color online) Chemical potential as a function of the gas temperature. Solid, dashed, and dotted lines correspond to $\tilde{\eta} = 0.01$, $\tilde{\eta} = 1$, and $\tilde{\eta} = 10$, respectively.}
\label{fig1}
\end{figure}

\begin{figure}
\epsfig{file=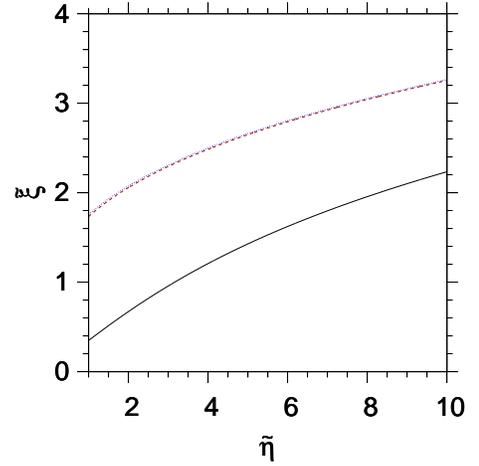,width=0.7\columnwidth}
\caption{(Color online) Chemical potential as a function of the gas density. Solid, dashed, and dotted lines correspond to $\tilde{\beta} = 1$, $\tilde{\beta} = 10$, and $\tilde{\beta} = 100$, respectively.}
\label{fig2}
\end{figure}

\begin{figure}
\epsfig{file=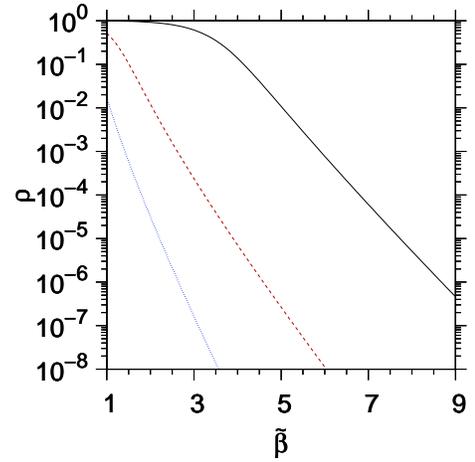,width=0.7\columnwidth}
\caption{(Color online) The ratio $\rho = N^+/N^-$ [cf. Eq. \eqref{cp4}] as a function of $\tilde{\beta}$. Solid, dashed, and dotted lines correspond to $\tilde{\eta} = 0.01$, $\tilde{\eta} = 1$, and $\tilde{\eta} = 10$, respectively.}
\label{fig3}
\end{figure}

\begin{figure}
\epsfig{file=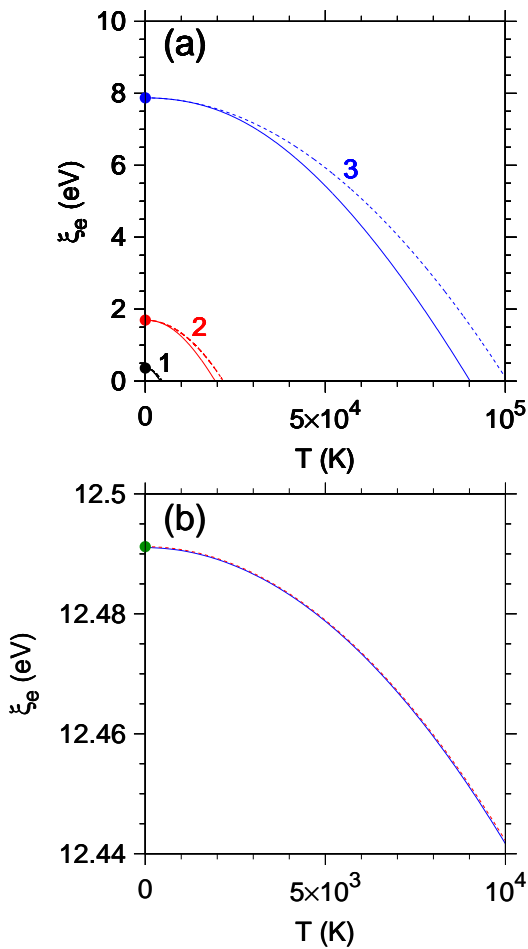,width=0.8\columnwidth}
\caption{(Color online) Chemical potential, measured with respect to the $mc^2$ rest energy [cf. Eq. \eqref{cp5}], as a function of the gas temperature for different values of the $\eta$ gas density. Solid and dashed lines corresponds to numerical results obtained from Eqs. \eqref{cp3} and \eqref{cp6}, respectively. The set of curves 1, 2, and 3 in panel (a) correspond to  $\eta=10^{21}$ cm$^{-3}$, $\eta=10^{22}$ cm$^{-3}$, and $\eta=10^{23}$ cm$^{-3}$, respectively. In panel (b), calculations were performed for $\eta$ corresponding to the electron density in silicon. The full dot in the left vertical axis corresponds to the Fermi energy obtained from Eq. \eqref{cp7}.}
\label{fig4}
\end{figure}

To compute the electromagnetic responses of the REG through Eqs.\eqref{e6}-\eqref{e9}, one needs to obtain the $\xi$ chemical potential which depends on the temperature and carrier density. The carrier density is $\eta=\Delta N/V$, where $\Delta N=N^--N^+$ is the difference between the $N^-$ number of particles and the $N^+$ number of antiparticles in the system. Then, one needs to solve the transcendental equation \cite{AragaoPRDS2016}
\be
\label{cp1}
\Delta N = N^- - N^+ = \sum_{\bvec{p}} g \, f (p,\beta,\xi),
\ee
where
\be
\label{cp2}
f (p,\beta,\xi) = \frac{1}{\e^{\beta \left ( \Omega_p - \xi \right )} + 1} - \frac{1}{\e^{\beta \left ( \Omega_p + \xi \right )} + 1}
\ee
is the distribution function accounting for the presence of both particles and antiparticles, $\Omega_p = \sqrt{p^2 c^2 + m^2 c^4}$ is the relativistic energy of a carrier with momentum $p$, and $g=2$ is the degeneracy factor of the electron gas. Eq. \eqref{cp1} reduces to

\be
\label{cp3}
\tilde \eta =\frac{\eta}{\eta_0} = \int_{0}^{+ \infty} dy \, y^2 \, {\cal F}_0 \left ( y,\tilde{\beta},\tilde{\xi} \right )
\ee
where $\eta_0 = g \,q_c^3/ (2 \pi^2) \approx 1.76 \times 10^{30}$ cm$^{-3}$ only depends on universal constants and may, therefore, be used as a natural unit to measure the $\eta$ effective carrier density of the REG. It should be noted [Eq. \eqref{eq11}] that ${\cal F}_0 \left ( y,\tilde{\beta},-\tilde{\xi} \right ) = -{\cal F}_0 \left ( y,\tilde{\beta},\tilde{\xi} \right )$, i.e, ${\cal F}_0 = {\cal F}_0 \left ( y,\tilde{\beta},\tilde{\xi} \right )$ is an odd function of the chemical potential. Eq. \eqref{cp3} implicitly defines the function $\tilde{\xi} = \tilde{\xi} \left ( \tilde{\beta},\tilde{\eta} \right )$. One sees that $\tilde{\xi} = 0$ leads to $\eta/\eta_0 = 0$, a case with corresponds to vacuum.

The chemical potential as a function of $\tilde{\beta}$ is displayed in Fig. \ref{fig1}. Calculations were performed for three different values of the density expressed in units of $\eta_0$. The numerical results suggest a weak temperature dependence of the chemical potential, if compared with $m c^2 \approx 0.511$ MeV, in the low-temperature limit ($\tilde{\beta} \rightarrow \infty$). Actually, the chemical potential exhibits variations of a few eV in the low-temperature limit, a fact which agrees with the non-relativistic theory of the electron gas (see below). 

The chemical potential is displayed in Fig. \ref{fig2} as a function of the density expressed in units of $\eta_0$. Numerical results were obtained for three different values of the gas temperature. The chemical potential is a growing monotonic function of $\tilde{\eta}$. One may note that the chemical potentials for $\tilde{\beta}=10$ and $\tilde{\beta}=100$ essentially coincide in the scale of the figure [cf. dashed and dotted lines in Fig. \ref{fig2}]. 

The ratio
\be
\label{cp4}
\rho = \frac{N^+}{N^-} = \frac{\int_{0}^{+ \infty} dy \, \frac{y^2}{\e^{\tilde{\beta} \left ( \sqrt{y^2 + 1} + \tilde{\xi} \right )} + 1}}{\int_{0}^{+ \infty} dy \, \frac{y^2}{\e^{\tilde{\beta} \left ( \sqrt{y^2 + 1} - \tilde{\xi} \right )} + 1}}.
\ee
as a function of $\tilde{\beta}$ is displayed in Fig.  \ref{fig3} for various values of the density. For a given temperature, it is apparent that the number of particles exceeds the number of antiparticles in all cases and the ratio $\rho = N^+/N^-$ decreases as the density of particles in the electron gas is increased, as expected. One may also note that $\rho \rightarrow 0$ as $\tilde{\beta} \rightarrow \infty$, since in the low-temperature limit one has $N^+ \ll N^-$. In other words, in the limit $\tilde{\beta} \rightarrow \infty$, the term corresponding to the occupation factor of antiparticles in the Fermi-Dirac distribution function  [cf. the second term in the RHS of Eq. \eqref{cp2}] may essentially  be neglected ($N^+ << N^-$) and the non-relativistic limit of the Fermi-Dirac distribution function is eventually recovered.

We have also explored the behavior of the chemical potential for density and temperature values appropriate for solid-state materials. In this respect, we have defined
\be
\label{cp5}
\xi_e (T) = \xi (T) - mc^2
\ee
as the non-relativistic chemical potential. According to the non-relativistic theory of the free-electron gas, it is well known that
\be
\label{cp6}
\xi_e (T) \approx E_F \left [ 1 - \frac{\pi^2}{12} \left ( \frac{T}{T_F} \right )^{\!\!\!2} \, \right ],
\ee
where
\be
\label{cp7}
E_F = \left ( \frac{6 \pi^2}{g} \eta \right )^{\!\!\!\frac{2}{3}} \frac{\hbar^2}{2m},
\ee
is the Fermi energy and $T_F = E_F / k_B$ is the Fermi temperature. The $\xi_e$  chemical potential is depicted in Fig. \ref{fig4} as a function of the gas temperature.  Calculations in Fig. \ref{fig4}(a) were performed for three different values of $\eta$ varying within the range exhibited by most of the solid-state materials. In Fig. \ref{fig4}(b) we have assumed $\eta \approx 2.0 \times 10^{23}$ cm$^{-3}$ corresponding to the electron density in silicon. The solid line corresponds to the result computed by combining Eqs. \eqref{cp3} and \eqref{cp5}, whereas the dashed line was obtained from Eq. \eqref{cp6}. The Fermi energy computed from the non-relativistic electron-gas model [cf. full dot in the vertical axis of Fig. \ref{fig4}(b)and Eq. \eqref{cp7}] essentially coincides with the numerical result obtained from Eqs. \eqref{cp3} and \eqref{cp5} in the limit $T \rightarrow 0$. Low-temperature results obtained from the non-relativistic model agree with those derived from the relativistic theory, as expected.

\subsection{Longitudinal plasmon modes of the REG}

\begin{figure}
\epsfig{file=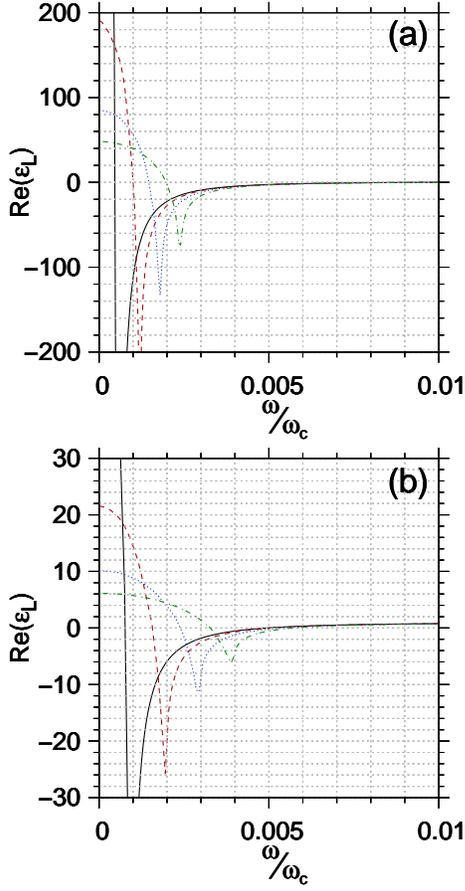,width=0.7\columnwidth}
\caption{(Color online) Real part of $\epsilon_L$ as a function of the $\omega$ frequency in units of the $\omega_c$ Compton frequency, for various values of the wave vector $\vert \vec{q} \vert$ expressed in units of the $q_c$ Compton wave vector, and for $\tilde{\eta}=0.01$. Results of panels (a) and (b) were obtained for $\tilde{\beta} = 1000$ and $\tilde{\beta} = 1$, respectively. Solid, dashed, dotted, and dot-dashed lines in panel (a) [(b)] correspond to $\tilde{q} = 0.002$, $\tilde{q} = 0.004$, $\tilde{q} = 0.006$, and $\tilde{q} = 0.008$, respectively ($\tilde{q} = 0.001$, $\tilde{q} = 0.002$, $\tilde{q} = 0.003$, and $\tilde{q} = 0.004$, respectively).}
\label{fig5}
\end{figure}

\begin{figure}
\epsfig{file=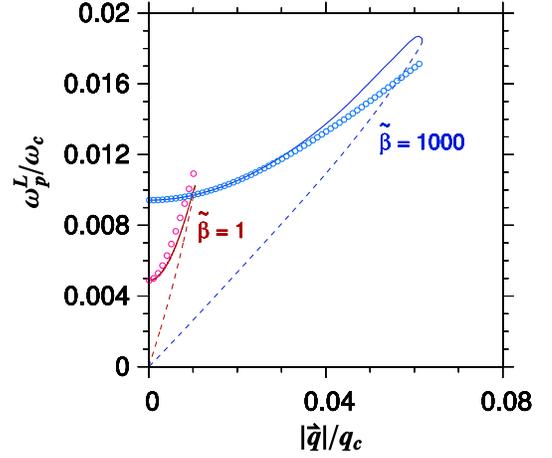,width=0.8\columnwidth}
\caption{(Color online) Upper ($\omega_{p}^L$ longitudinal plasmon frequency) and lower frequency zeros (solid and dashed lines, respectively) of the real part of $\epsilon_L$  as functions of the $\vert \vec{q} \vert$ wave vector. Calculations were performed for $\tilde{\beta} = 1$ and $\tilde{\beta} = 1000$, by taking $\tilde{\eta} = 0.01$. Open circles correspond to the analytical results obtained from Eq. \eqref{lpe10}. Both $\omega$ and $\vert \vec{q} \vert$ are expressed in units of $\omega_c$ and $q_c$, respectively.}
\label{fig6}
\end{figure}

\begin{figure}
\epsfig{file=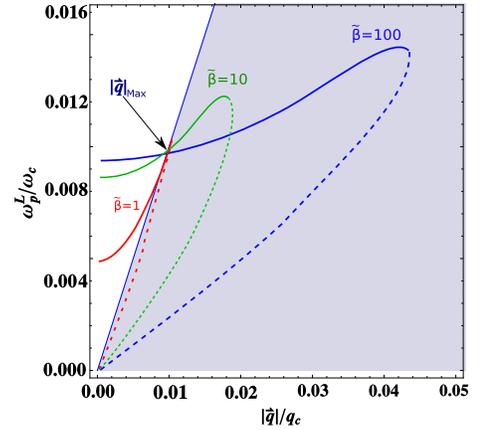,width=0.7\columnwidth}
\caption{(Color online) Upper ($\omega_p^L$  plasmon frequency) and lower frequency zeroes of $\epsilon_L$  (solid and dashed lines, respectively) as functions of the $\vert \vec{q} \vert$ wave vector. Calculations were performed for $\tilde{\beta} = 100$, $\tilde{\beta} = 10$ and $\tilde{\beta} = 1$ by taking $\tilde{\eta} = 0.01$. The shaded area corresponds to the region where particle-antiparticle excitations occur. Note the maximum value of the wave vector ($\tilde{q}_{max}$) beyond which the longitudinal plasmon decays into particle-antiparticle pairs. Both the frequency and wave vector are given in units of $\omega_c$ and $q_c$, respectively.}
\label{fig7}
\end{figure}

\begin{figure}
\epsfig{file=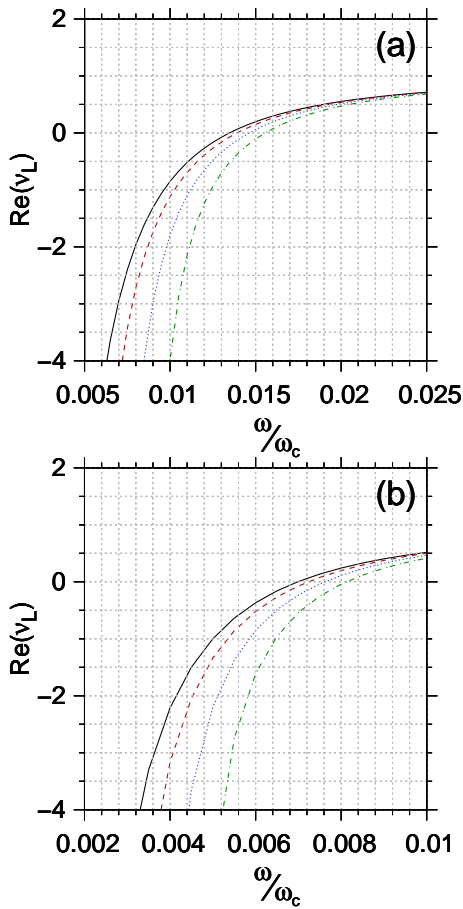,width=0.7\columnwidth}
\caption{(Color online) Real parts of $\nu_L$ as a function of the $\omega$ frequency in units of the $\omega_c$ Compton frequency, for various values of the wave vector $\vert \vec{q} \vert$ given in units of the $q_c$ Compton wave vector. Calculations were performed for $\tilde{\eta} = 0.01$. Results of panels (a) [(b)] were obtained for $\tilde{\beta} = 1000$ ($\tilde{\beta} = 1$). Solid, dashed, dotted, and dot-dashed lines in panels (a) [(b)] correspond to $\tilde{q} = 0.002$, $\tilde{q} = 0.004$,  $\tilde{q} = 0.006$, and $\tilde{q} = 0.008$, respectively ($\tilde{q} = 0.001$, $\tilde{q} = 0.002$,  $\tilde{q} = 0.003$, and $\tilde{q} = 0.004$, respectively).}
\label{fig8}
\end{figure}

Now we focus our attention on the $ \epsilon_{ij} $ electric permittivity tensor. According to Eq. \eqref{long1} one has

\be
\epsilon_L = \epsilon+\epsilon'= 1+{\cal C}^\ast + \left(1-\frac{\tilde{\omega}^2}{\tilde{q}^2}\right) {\cal B}^\ast,
\ee

The real part of $\epsilon_L$ is depicted in Fig. \ref{fig5} as a function of $\omega$ in units of $\omega_c$. Results were computed for $\tilde{\eta}=0.01$ and different values of $\vert \vec{q} \vert$ expressed in units of $q_c$. In Figs. \ref{fig5}(a) and \ref{fig5}(b) we have set $\tilde{\beta} = 1000$ and $\tilde{\beta} = 1$, respectively. The real part of the longitudinal electric permittivity exhibits two zeros for a given value of $\tilde{q}$ at a given temperature. The lower zero lies within a region where $\mathrm{Im} [\epsilon_L ] \neq 0$. Therefore, in spite of the fact that $\mathrm{Re} [ \epsilon_L ] = 0$ in this case, such a zero cannot be considered as a plasmon frequency. In the vicinity of the upper zero, on the other hand, one may have $\mathrm{Im} [ \epsilon_L ] = 0$ (see discussion below), so that it corresponds to the $\omega_{p}^L$ longitudinal plasmon frequency. We then show in Fig. \ref{fig6} the upper (solid lines) and lower (dashed lines) frequency zeros of the real part of $\epsilon_L$ as functions of $\tilde{q}$. Numerical calculations were performed for $\tilde{\eta} = 0.01$ and two different values of $\tilde{\beta}$. 

For sufficiently small values of the $\tilde{q}$ wave vector, the longitudinal-plasmon dispersion relation may be approximately described by the expression
\be
\label{lpe4}
(\tilde{\omega}_{p}^L)^2 = (\tilde{\omega}_{0p}^L)^2 + \frac{v_T^2}{c^2} \tilde{q}^2,
\ee
where the frequencies and wave vector are given in units of $\omega_c$ and $q_c$, respectively. In the above expression $\tilde{\omega}_{0p}^L$ is the temperature-dependent longitudinal plasmon frequency in the long-wavelength limit and $v_T$ plays the role of the carrier thermal speed, except for a numeric factor. The low-temperature non-relativistic case $v_T^2 = 3/(m \beta)$ leads to the well-known Bohm-Gross dispersion relation \cite{Bohm}. To estimate the carrier thermal speed corresponding to the REG, we have computed the average of $p^2$ from the expression
\be
\label{lpe5}
\langle p^2 \rangle = \frac{\sum \limits_{\bvec{p}} g \, p^2 \, f (p,\beta,\xi)}{\sum \limits_{\bvec{p}} g \, f (p,\beta,\xi)},
\ee
which reduces to
\be
\label{lpe6}
\langle p^2 \rangle = m^2 c^2 {\cal K} (\tilde{\beta},\tilde{\xi}),
\ee
where
\be
\label{lpe7}
{\cal K} (\tilde{\beta},\tilde{\xi}) = \frac{1}{\tilde{\eta}}\int \limits_{0}^{+ \infty} dy \, y^4 \, {\cal F}_0 \! \left ( y,\tilde{\beta},\tilde{\xi} \right ).
\ee
The $v_T$ carrier thermal speed is given by
\be
\label{lpe8}
\langle p^2 \rangle = \frac{m^2}{1 - \frac{v_T^2}{c^2}} \, v_T^2,
\ee
which results in
\be
\label{lpe9}
v_T^2 = c^2 \frac{{\cal K} (\tilde{\beta},\tilde{\xi})}{1+{\cal K} (\tilde{\beta},\tilde{\xi})}.
\ee
Then Eq. \eqref{lpe4} becomes
\be
\label{lpe10}
(\tilde{\omega}_{p}^L)^2 = (\tilde{\omega}_{0p}^L)^2 + \frac{{\cal K} (\tilde{\beta},\tilde{\xi})}{1+{\cal K} (\tilde{\beta},\tilde{\xi})} \, \tilde{q}^2.
\ee
The above equation is a generalization of the Bohm-Gross dispersion relation to the case of a REG at finite temperatures and is valid in the limit $\tilde{q} \ll \tilde{q}_{max}$, i.e, far from the region where particle-antiparticle excitations occur. In condensed matter, that corresponds to electron-hole pairs whereas in a relativistic gas that corresponds to electron-positron pairs. Theoretical results obtained from Eq. \eqref{lpe10} are displayed in Fig. \ref{fig6} as open circles, for both $\tilde{\beta} = 1$ and $\tilde{\beta}=1000$, and for $\tilde{\eta} = 0.01$. In the long-wavelength regime, the agreement with the curve from Eq. \eqref{eplasmon} is quite good. 

We display in Fig. \ref{fig7} the upper (solid lines) and lower (dashed lines) frequency zeros of the real part of $\epsilon_L$ as functions of $\tilde{q} = \vert \vec{q} \vert / q_c$. Numerical results were computed for $\tilde{\eta} = 0.01$ and different values of $\tilde{\beta}$. Fig. \ref{fig7} clearly indicates that there is a  maximum value of the wave vector ($\tilde{q}_{max}$) beyond which the longitudinal plasmon decays into particle-antiparticle pairs. 

\subsection{Transverse plasmon modes of the REG}

\begin{figure}
\epsfig{file=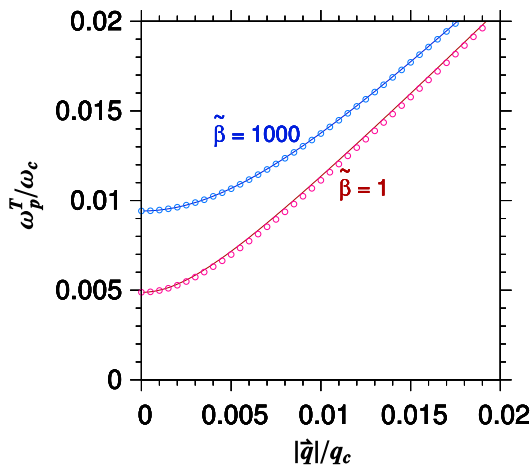,width=0.8\columnwidth}
\caption{(Color online) Transverse plasmon frequency (zero of $\nu_L+1 = 0$) as a function of the $\vert \vec{q} \vert$ wave vector. The $\omega_{p}^T$ transversal plasmon frequency and the $\vert \vec{q} \vert$ wave vector are given in units of the $\omega_c$ Compton frequency and $q_c$  Compton wave vector, respectively. Solid lines correspond to theoretical results obtained from $\nu_L+1 = 0$. Calculations were performed for $\tilde{\beta} = 1$ and $\tilde{\beta} = 1000$, by taking $\tilde{\eta}$ = 0.01. Open circles correspond to the analytical results obtained from Eq. \eqref{tp4}.} 
\label{fig9}
\end{figure}

The condition $\nu_L = -1$ [cf. \eqref{mplasmon}] leads to the $\omega_{p}^T$ frequency of the REG transverse plasmon modes. We display in Fig. \ref{fig8} the real part of $\nu_L$ as a function of the $\omega$ frequency in units of $\omega_c$, obtained for $\tilde{\eta} = 0.01$ and different values of the wave vector $\vert \vec{q} \vert$ in units of $q_c$. Results depicted in Figs. \ref{fig8}(a) and \ref{fig8}(b) where computed for $\tilde{\beta} = 1000$ and $\tilde{\beta} = 1$, respectively. We would like to stress that numerical results (not shown here) indicate that $\mathrm{Im} \left [ \nu_L \right ] = 0$ within the respective frequency ranges considered in both Fig. \ref{fig8}(a) and \ref{fig8}(b). Therefore, in the present cases, the transversal plasmon frequencies may be obtained by solving the transcendental equation $\mathrm{Re} \left [ \nu_L \right ] = -1$.

The $\omega_{p}^T$ transverse plasmon frequency is displayed in Fig. \ref{fig9} as a function of the $\tilde{q} = \vert \vec{q} \vert/q_c$ wave vector for different values of the gas temperature. Calculations were performed for $\tilde{\eta}$ = 0.01. Solid lines correspond to the numerical results obtained from $\nu_L = -1$ [cf. Eq.\eqref{mplasmon}]. Such results indicate the existence of one single positive zero of $\nu_L+1$ for a given value of $\tilde{q}$ at a given gas temperature. Therefore, the dispersion exhibits only a single branch. We also note that, at a given temperature, the wave vector dependence of the resonance frequency may be approximated by
\be
\label{tp4}
(\tilde{\omega}_{p}^T)^2 (q) = (\tilde{\omega}_{0p}^T)^2 + \tilde{q}^2,
\ee
which fits quite well the numerical results obtained from Eq. \eqref{mplasmon}. In the above equation $\tilde{\omega}_{0p}^T$ is the temperature-dependent transverse plasmon frequency numerically obtained from Eq. \eqref{mplasmon} in the limit $\tilde{q} \rightarrow 0$. Results obtained from Eq. \eqref{tp4} are displayed in Fig. \ref{fig9} as open circles.

\subsection{Decays and responses of the REG}

\begin{figure}
\epsfig{file=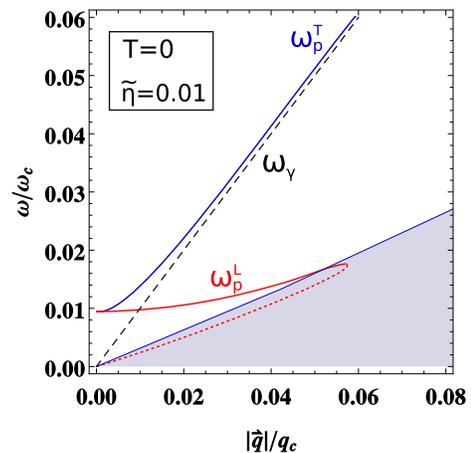,width=0.7\columnwidth}
\caption{(Color online) The dispersion curves for transverse and longitudinal plasmon modes at $T=0$ and $\tilde{\eta} = 0.01$ [cf. Eqs. \eqref{eplasmon} and \eqref{mplasmon}]. In the shaded area, $\mathrm{Im} \left [ \epsilon_L \right ] \ne 0$, indicating decay of the longitudinal mode. The dashed line is discarded as a solution for the longitudinal dispersion as it lies entirely in the region of nonzero imaginary part of $\epsilon_L$. We have also shown the dispersion (dashed line) for the photon mode $\tilde\omega_\gamma = \tilde q$ [see Eq. \eqref{photon}].}
\label{fig10}
\end{figure}

We display in Fig. \ref{fig10} the dispersion curves for the transverse and longitudinal plasmon modes at $T=0$ [cf. Eqs. \eqref{eplasmon} and \eqref{mplasmon}]. The shaded area corresponds to the region where the excitation of particle-antiparticle pairs occurs. The dashed line is discarded as a solution for the longitudinal plasmon dispersion as it lies entirely in the region of nonzero imaginary part of $\epsilon_L$. We have also shown the dispersion for the photon mode $\tilde\omega_\gamma = \tilde q$ [cf. \eqref{photon}]. Although not shown in the figure, the straight dotted line for the photon dispersion will eventually reach the upper region where the excitation of electron-positron pairs will take place \cite{AragaoPRDS2016}.

\begin{figure}
\epsfig{file=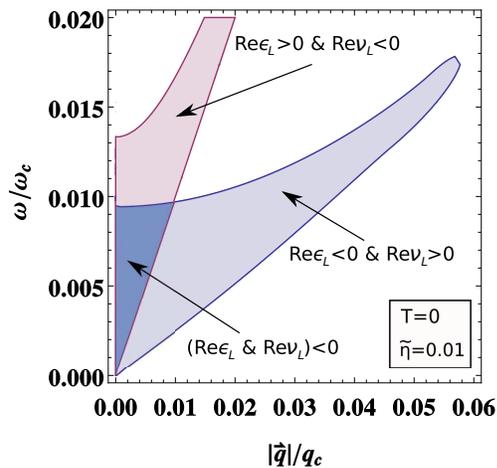,width=0.75\columnwidth}
\caption{(Color online) Regions of the $(\tilde{q},\tilde{\omega})$ plane according to the signs of the real parts of $\epsilon_L$ and $\nu_L$. Outside the shaded regions the real parts of $\epsilon_L$ and $\nu_L$ are positive. Results were obtained for $T=0$ and $\tilde{\eta}=0.01$.}
\label{fig11}
\end{figure}

Finally, we display in Fig. \ref{fig11} the different relevant regions of the $(\tilde{q},\tilde{\omega})$ plane where the real parts of $\epsilon_L$ and $\nu_L$ have different signs \cite{AragaoSubmit}.{We would like to stress there is a region where both $\epsilon_L$ and $\nu_L$ are simultaneously negative, indicating that the REG exhibits a behavior has not been experimentally observed in natural materials. This fact was previously remarked by one of the authors \cite{AragaoPRDS2016}, as mentioned. It is important to note that, in contrast to the non-relativistic case where the refractive index is defined as $n = \sqrt{\epsilon} \sqrt{\mu}$, simultaneous negative values of $\epsilon_L$ and $\nu_L$ observed in the present relativistic case do not imply negative refraction. 

A figure similar to Fig. \ref{fig11} can be obtained using the values of densities and temperatures encountered in astrophysical scenarios, as in a superdense electron-plasma (e-p) in Gamma-ray bursts (GRBs) \cite{berezhiani2015, Aksenov2010}, where the e-p density is in the range of $\eta=(10^{30} - 10^{37})\text{cm}^{-3}$.
 
According to ref.\cite{berezhiani2015}, in Condensed Matter, e-p plasmas will eventually be produced in the laboratory with laser systems. Laser pulses with focal densities $I=10^{22} \text{W cm}^{-2}$ incident on material targets could lead to e-p plasmas with the densities in the range of $(10^{23}-10^{28})\text{cm}^{-3}$. We have used the upper limit of that density range in our calculations.	
	
A study of the refractive index of  the REG will be the subject of a further investigation, in which we intend to generalize the results of Lepine and Lakhatakia \cite{Lakhtakia2002,Lakhtakia2004}.

\section{Conclusions}
\label{conclusions}

Summing up, we have presented a theoretical study of the EM propagation and responses of a REG for various temperatures and carrier densities. Using linear response and RPA, we have identified the propagation modes and their dispersion relations from the QED propagators as well as from Maxwell's equations with the added input of the constitutive relations obtained from the QED responses. We found a longitudinal plasmon mode, two transverse plasmon modes, and a photonic mode which propagates with the speed of light in vacuum, i.e., for which the medium is transparent thanks to the specific form of its relativistic electromagnetic responses. In deriving dispersion relations, we were able to identify stable solutions and regions of instability where the plasmon modes decay. Finally, we have also identified the regions in the $(\vert \vec q \vert,\omega)$ plane where the longitudinal permittivity $\epsilon_L $ and longitudinal inverse permeability $\nu_L$ are  both simultaneously negative. 

\acknowledgments

The authors would like to thank the Scientific Colombian Agency CODI - University of Antioquia, and Brazilian Agencies CNPq, CAPES, FAPESP, and FAEPEX/UNICAMP for partial financial support.

\end{document}